\documentclass[nofootinbib,aps,prd,nobibnotes,twocolumn,superscriptaddress,bibliography]{revtex4-2}

\usepackage{amsfonts}
\usepackage{mathrsfs}
\usepackage{amsmath}
\usepackage{color}
\usepackage{graphicx}
\usepackage{bm}
\usepackage{amssymb}
\usepackage{xspace}
\usepackage{epstopdf}
\usepackage{multirow}
\usepackage{braket}
\usepackage{soul}
\usepackage{xcolor}
\usepackage{tabularx}
\newcolumntype{Y}{>{\centering\arraybackslash}X}

\usepackage[colorlinks=true, letterpaper=true, pdfstartview=FitV, linkcolor=blue, citecolor=blue, urlcolor=blue]{hyperref}

\makeatother
\begin{document}
	\title{Hilbert band complexes and their applications}

\author{Zeying Zhang}\email{zzy@mail.buct.edu.cn}
\affiliation{College of Mathematics and Physics, Beijing University of Chemical Technology, Beijing 100029, China}

\author{Y. X. Zhao}
\affiliation{Department of Physics and HKU-UCAS Joint Institute for Theoretical
	and Computational Physics at Hong Kong, The University of Hong Kong,
	Pokfulam Road, Hong Kong, China}

\author{Yugui Yao}
\affiliation{Centre for Quantum Physics, Key Laboratory of Advanced Optoelectronic Quantum Architecture and Measurement (MOE), Beijing Institute of Technology, Beijing 100081, China}

\author{Shengyuan A. Yang}
\affiliation{Research Laboratory for Quantum Materials, IAPME, FST, University of Macau, Macau, China}

\begin{abstract}
The study of band connectivity is a fundamental problem in condensed matter physics. Here, we develop a new method for analyzing band connectivity, which completely solves the outstanding questions of the reducibility and decomposition of band complexes. By translating the symmetry conditions into a set of band balance equations, we show that all possible band structure solutions can be described by a positive affine monoid structure, which has a unique minimal set of generators, called Hilbert basis. We show that  Hilbert basis completely determine whether a band complex is reducible and how it can be decomposed. The band complexes corresponding to Hilbert basis vectors, termed as Hilbert band complexes (HBCs), can be regarded as elementary building blocks of band structures. We develop algorithms to construct HBCs, analyze their graph features, and merge them into large complexes. We find some interesting examples, such as HBCs corresponding to complete bipartite graphs, and complexes which can grow without bound by successively merging a HBC.

\end{abstract}
\maketitle

\section{Introduction}

\begin{figure}[t]
	\includegraphics[width=\linewidth]{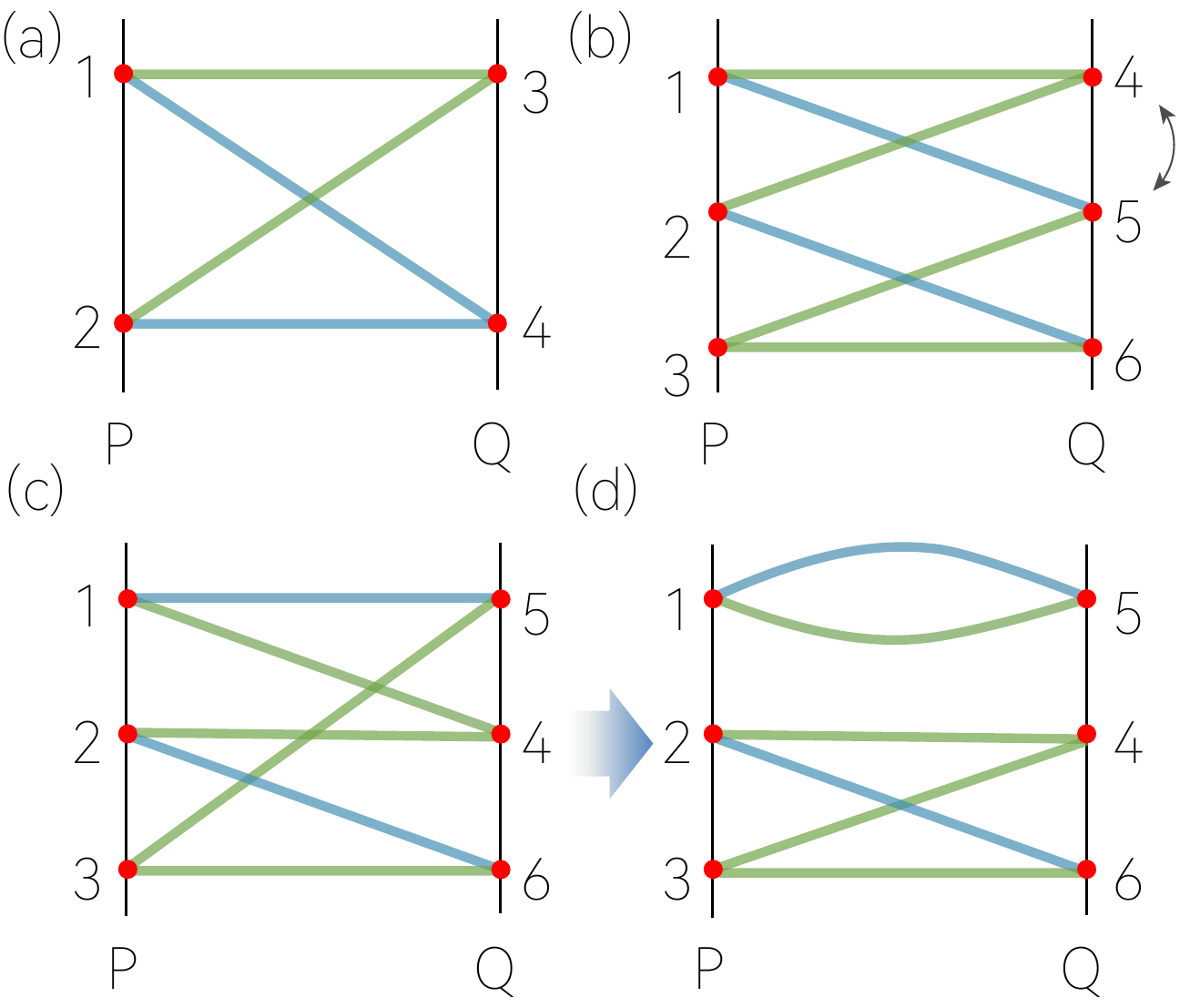}
	\caption{(a) Hourglass type band complex. The four vertices (corresponding to twofold degenerate states at $P$ and $Q$) are denoted by red dots. The colors of the edges (corresponding to band curves on path $P$-$Q$) reflect their different symmetry characters.
 (b) An example of a reducible band complex. By switching the energy ordering of vertices 4 and 5 (indicated by the arrow), the complex becomes that in (c). However, (c) is unstable, because the edges with the same color cross on the path (indicated by the arrows). These crossing points will be gapped, and the complex will be decomposed into two sub-complexes, as shown in (d).}
	\label{fig:intro}
\end{figure}

Band theory is a fundamental pillar of condensed matter physics \cite{ashcroft_solid_1976}. The distribution of energy bands and band gaps along energy axis provides us basic understanding of metals and insulators 
Recent research on topological states of matter further directs our attention towards more detailed band connectivity properties in energy-momentum space \cite{chiu_classification_2016, armitage_weyl_2018,  yan_topological_2017,  bradlyn_topological_2017, vergniory_graph_2017}. For example, it was found that certain space group symmetries can enforce an hourglass type band structure on a high-symmetry path of Brillouin zone (BZ) \cite{young_dirac_2015, wang_hourglass_2016, bzdusek_nodal-chain_2016,wang_hourglass_2017, ma_experimental_2017, fu_hourglasslike_2018, wu_exhaustive_2020}, as illustrated in Fig.~\ref{fig:intro}(a). Clearly, this band pattern forms a connected graph. Here, one defines the \emph{vertices} of the graph as corresponding to the (possibly degenerate) states at high-symmetry points [$P$ and $Q$ in Fig.~\ref{fig:intro}(a)]. For the hourglass band pattern, there are four vertices, marked by the red colored dots in Fig.~\ref{fig:intro}(a). The \emph{edges} of the graph correspond to the band curves between the vertices. The graph is connected if one can traverse all the vertices through the edges. Such a connected band pattern is called a \emph{band complex}.

A band complex may contain many bands. For example, in Ref.~\cite{li_upper_2023}, a band complex was constructed which may contain arbitrarily large number $N_C=4n$ ($n\in \mathbb{N}$) of bands. Generally, such large band complexes are \emph{reducible}, meaning that by switching the energy ordering of some of its vertices, the complex can be decoupled into disconnected components. For instance, consider the band complex in Fig.~\ref{fig:intro}(b). The colors of the band curves indicate their symmetry characters on the path $P$-$Q$. By switching the energy ordering of vertices $4$ and $5$, one can easily see that the complex would be decomposed into two smaller complexes, as shown in Fig.~\ref{fig:intro}(c) and \ref{fig:intro}(d). In comparison, the hourglass complex in Fig.~\ref{fig:intro}(a) is \emph{irreducible}: it remains connected no matter how the vertices are switched in energy.

From the above observation, there is a natural and fundamental question regarding band complexes, namely, \emph{given a band complex, how can we determine whether it is reducible or not?} Related to this question, one may further ask: \emph{If it is reducible, can we determine its decomposition?}

In this work, we provide a complete answer to these questions. Our solution is to convert the question about band complexes into an algebraic problem with an affine monoid structure. We show that the basis of this affine monoid determine a set of
irreducible band complexes, termed as the Hilbert band complexes (HBCs). We classify all possible HBCs on a high-symmetry path for the 230 nonmagnetic space groups.
With the help of HBCs, we can readily determine whether a complex is reducible and how it can be decomposed. Interestingly, we show that for certain band complexes, the decomposition is not unique, i.e., it may be decomposed into two (or more) different sets of HBCs. We also propose an algorithm to construct larger band complexes from HBCs. Along this way, we find some interesting examples, where certain band complexes can `grow' without bound by consecutively adding in some simple HBCs. These developments provide a powerful method for analyzing band structures and deepen our understanding of the fundamental aspects of band theory.

\section{band balance equations}

To begin with, consider a path $U$ connecting two high-symmetry points, denoted as $P$ and $Q$, in a BZ corresponding to some space group.  The vertices at $P$ ($Q$) are each labeled by the irreducible representations (IRRs) of the little group $G_P$ ($G_Q$) at this point. The band curves along $U$ are labeled by the IRRs of the little group $G_U$ of this path.

Because $G_U$ must be a subgroup of $G_P$ and $G_Q$, a degenerate vertex often split and emit multiple edges due to the symmetry reduction when going from $P$ ($Q$) to generic points on $U$. This splitting is governed by the so-called compatibility relations \cite{bouckaert_theory_1936, bradley_mathematical_2009}. These relations take the form of
\begin{equation}
  \mathcal{P}_i\downarrow U=\bigoplus_{j=1}^s a^{PU}_{ji}\mathcal U_j,
\end{equation}
where $\mathcal{P}_i$ $(i=1,\cdots,m)$ are the IRRs of $G_P$, $\mathcal{U}_j$ $(j=1,\cdots,s)$ are the IRRs of $G_U$,
the coefficients $a^{PU}_{ji}\in \mathbb{Z}_{\geq 0}$ are nonnegative integers, and integers $m$ and $s$ are finite since we are dealing with finite groups. Similarly, for the other end at $Q$, we have
\begin{equation}
  \mathcal{Q}_i\downarrow U=\bigoplus_{j=1}^s a^{QU}_{ji}\mathcal U_j,
\end{equation}
where $\mathcal{Q}_i$ $(i=1,\cdots,n)$ are the IRRs of $G_Q$.

Now, consider a band structure on this path. It may be a single band complex, or it may contain multiple band complexes (i.e., multiple disconnected components). Suppose there are $c_i^P$ number of vertices at point $P$ with IRR $\mathcal{P}_i$ $(i=1,\cdots,m)$ and $c_\ell^Q$ number of vertices at point $Q$ with IRR $\mathcal{Q}_\ell$ $(\ell=1,\cdots,n)$. By the compatibility relations, the vertices at $P$ and the vertices at $Q$ cannot be arbitrary; they must be balanced to produce the same {\color{black}multiplicities} for each IRR $\mathcal{U}_j$ on path $U$. This gives us the basic condition:
\begin{equation}
  \bigoplus_{i=1}^m c_i^P (\mathcal{P}_i\downarrow U)=\bigoplus_{\ell=1}^n c_\ell^Q (\mathcal{Q}_\ell\downarrow U).
\end{equation}
This condition is actually a set of $s$ coupled linear equations, one for each $\mathcal{U}_j$. Explicitly, we have
for each $j$ ($=0,1,\cdots,s$):
\begin{equation}\label{BB}
  \sum_{i=1}^m a_{ji}^{PU} c_i^P-\sum_{\ell=1}^n a_{j\ell}^{QU}c_\ell^Q=0.
\end{equation}
These equations constrain the possible values of the numbers $c_i^P$ and $c_\ell^Q$. We will refer to these equations as band balance equations.

\section{Hilbert basis}

By definition, $c_i^P$ and $c_\ell^Q$ are nonnegative integers.   We may put $A^P=[a_{ji}^{PU}]$ as a $s\times m$ matrix and $A^Q=[a_{j\ell}^{QU}]$ as a $s\times n$ matrix. Define an $s\times (m+n)$ matrix
\begin{equation}
  \mathbf A=[A^P, -A^Q],
\end{equation}
and define a column vector $\mathbf c=(c_1^P,\cdots, c_m^P, c_1^Q,\cdots c_n^Q)^T$.
Then the band balance equations take the form of
\begin{equation}\label{66}
  \mathbf A\mathbf c=0.
\end{equation}
{\color{black} If we consider the solution of the band balance equations in the domain of $\mathbb{Q}$ or $\mathbb{R}$, then the problem would be greatly simplified.} Its solution is the kernel of $\mathbf A$, which has a nice vector space structure and can be readily obtained
by standard approaches, such as Gaussian elimination method.

From the above construction, a band complex $B$ on path $P$-$Q$ naturally defines a $\textbf c$ vector, which satisfies Eq.~(\ref{66}). The {\color{black}dimension} of $\mathbf c$ is the number of IRRs at $P$ and $Q$. It is termed as the $\textbf c$-vector of the band complex $B$, denoted by $\textbf c(B)$. Clearly, such $\textbf c(B)$ is a solution of Eq.~(\ref{66}). However, solutions of Eq.~(\ref{66}) also include `unphysical' ones for which some entries of the vector are not nonnegative integers.

When we restore the requirement that $c_i^P,c_\ell^Q\in \mathbb{Z}_{\geq 0}$, the valid solutions will correspond to
the set
\begin{equation}\label{setS}
  S=\text{ker} \mathbf A \cap \mathbb{Z}_{\geq 0}^{m+n}.
\end{equation}
Evidently, $S$ does not have a vector space structure, and it is not even a group, since a nonzero element does not have an inverse. Nevertheless, according to {Gordan's lemma}, $S$ is a good example of a positive affine monoid \cite{gubeladze_polytopes_2009}. A  {\color{black} commutative} monoid is a set with a {\color{black} binary multiplication} (addition here), which is closed, {\color{black}is commutative}, is associative, and has identity element (which is the zero vector). `Positive' means only the zero element has inverse. `Affine' indicates that the monoid is finitely generated, namely, the whole set can be generated by a finite number of generators. More importantly, among all possible generators, there exits a \emph{unique} set of generators which are minimal in number. Such a set is called a Hilbert basis \cite{gubeladze_polytopes_2009}.
Each element of Hilbert basis cannot be generated by other elements in $S$, i.e., they are irredundant. Denoting the Hilbert basis for $S$ by {\color{black}a set of Hilbert basis vectors} $\{\mathbf v^i\}$ ($i=1,\cdots, d$), then we have
\begin{equation}
  S=\Big\{q_1 \mathbf v^1+q_2 \mathbf v^2+\cdots+q_d \mathbf v^d\ |\ q_i\in \mathbb{Z}_{\geq 0}\Big\}.
\end{equation}
The {\color{black}cardinality} of Hilbert basis must be greater than or equal to the dimension of $\text{ker} \mathbf A$, namely,
$d\geq \text{dim}\, \text{ker} \mathbf A$.

Before proceeding, two points must be emphasized. First, for a given positive affine monoid $S$, its Hilbert basis is unique, so without any ambiguity, we may denote the Hilbert basis by $\text{Hilb}(S)$. This is in sharp contrast to vector spaces, whose bases are not unique. Second, for a vector space $V$, the decomposition of any element of $V$ onto a set of chosen basis is unique.
In contrast, the decomposition of an element of $S$ by Hilbert basis may not be unique. In other words, there may be more than one way to generate a particular element by the Hilbert basis.

\begin{figure}[t]
	\includegraphics[width=\linewidth]{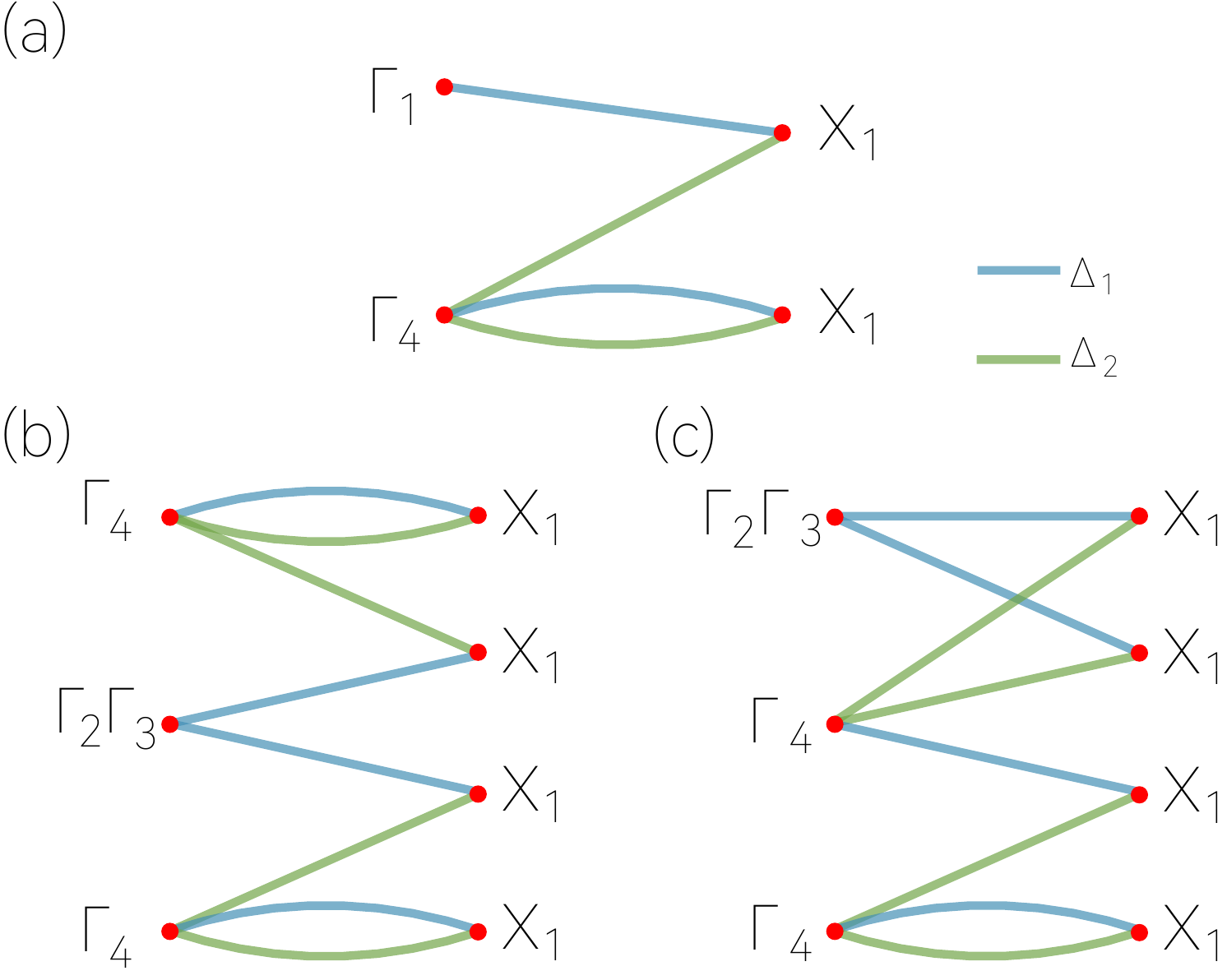}
	\caption{Hilbert band complexes on the $\Delta$ path of space group No.~198. The symbols at vertices indicate the IRRs of the states at $\Gamma$ and $X$. The blue and green colors correspond respectively to the $\Delta_1$ and $\Delta_2$ IRRs of the band curves on the $\Gamma$-$X$ ($\Delta$) path.
(a) Hilbert band complex corresponding to $\mathbf v^1$ in Eq.~(\ref{15}). (b, c) Two Hilbert band complexes corresponding to $\mathbf v^2$ in Eq.~(\ref{15}).
Viewed as graphs, (b) and (c) are topologically distinct, but they correspond to the same Hilbert basis vector.}
	\label{fig:198}
\end{figure}

{

To illustrate the idea above, let's consider a simple example. Take space group $P2_1 3$ (No.~198). Consider the path
$\Delta: (0,u,0)$ which connects the two high-symmetry points $\Gamma:(0,0,0)$ and $X:(0,1/2,0)$. In the spinless case (i.e., considering the single-valued representations), there are three IRRs at $\Gamma$: $\Gamma_1\ (1)$, $\Gamma_2\Gamma_3\ (2)$, and $\Gamma_4\ (3)$ \cite{bradley_mathematical_2009, liu_spacegroupirep_2021, liu_msgcorep_2023}. Here, the number in the parenthesis after the IRR symbol gives the dimension of the IRR.
At $X$ point, there is only one IRR: $X_1\ (2)$. On the path $\Delta$, there are two 1d IRRs: $\Delta_1\ (1)$ and $\Delta_2\ (1)$. The compatibility relations are given by
\begin{align}\label{99}
  \Gamma_1\downarrow \Delta&=\Delta_1,\\
  \Gamma_2\Gamma_3\downarrow \Delta&=2\Delta_1,\\
  \Gamma_4\downarrow \Delta&=\Delta_1\oplus 2\Delta_2,\\
  X_1\downarrow \Delta&=\Delta_1\oplus \Delta_2.
\end{align}
Then, the $\textbf A$ matrix in this case is
\begin{equation}
  \textbf A=\left(
              \begin{array}{cccc}
                1 & 2 & 1 & -1 \\
                0 & 0 & 2 & -1 \\
              \end{array}
            \right).
\end{equation}
The kernel of $\textbf A$ is a two-dimensional subspace. A choice of its basis may be taken as
\begin{equation}
 \mathbf e^1=(-2, 1, 0, 0)^T,\qquad \mathbf e^2=(1, 0, 1, 2)^T.
\end{equation}
However, $\mathbf e^1$ is clearly not an element of $S=\text{ker} \mathbf A \cap \mathbb{Z}_{\geq 0}^4$, since its first entry is a negative integer. Instead, one can check that all elements of $S$ can be generated by the following unique Hilbert basis:
\begin{equation}\label{15}
 \mathbf v^1=(0, 1, 2, 4)^T,\qquad \mathbf v^2=(1, 0, 1, 2)^T.
\end{equation}
The four numbers in each basis vector correpond to the four IRRs $\Gamma_1$, $\Gamma_2\Gamma_3$, $\Gamma_4$, and $X_1$.

In this particular example, the number of elements in $\text{Hilb}(S)$ happens to be the same as $\text{dim}\, \text{ker} \mathbf A$. This needs not be the case in general situations. $\text{dim}\, \text{ker} \mathbf A$ is the lower bound for the size of  $\text{Hilb}(S)$. In simple cases, the Hilbert basis may be determined by observation.
However, for more complicated cases, the determination is not easy. Finding the Hilbert basis of an affine monoid is
a classical problem in algebraic geometry. Mathematicians have developed several computation algorithms to achieve this.
Generally, the procedure has two main steps \cite{gubeladze_polytopes_2009}.
First, one finds a set (generally not minimal) of  generators for the given affine monoid. This step can be done by identifying the generators of rational cones\footnote{\color{black}Here, a cone $C$ is the intersection of finitely linear closed halfspaces over $\mathbb{R}$, a rational cone $R$ is submonoid of $C$ such that $R=C\cap \mathbb{Q}$. }, whose intersection forms the affine monoid.
Second, one reduces this set to a Hilbert basis. A possible approach (may not be the fastest one) is to adopt the Fourier-Motzkin elimination method \cite{gubeladze_polytopes_2009}.
Several software packages, such as
\textsf{Normaliz} \cite{bruns_normaliz_2010}, \textsf{4ti2} \cite{4ti2_team_4ti2software_nodate} and \textsf{LattE} \cite{the_latte_team_latte_nodate}, have been developed for the task of solving Hilbert basis. In this work, we use \textsf{Normaliz} to calculate the Hilbert basis in our problem.

\section{Hilbert band complex}
\label{sec:graph}

Regarding band connectivity, the significance of the Hilbert basis obtained above lies in the following result.

\textbf{Proposition 1.} A band complex $B$ on path $P$-$Q$ is irreducible \emph{if and only if} its
$\mathbf c$-vector $\mathbf c(B)$ is a element of $\text{Hilb}(S)$.

\emph{Proof.} Let's first prove the `if' part, namely, if $\mathbf c(B)$ is a Hilbert basis vector, then this band complex $B$ is irreducible. This can be proved by contradiction. Suppose $B$ were instead reducible, meaning that by changing the energy ordering of its vertices, $B$ can be decoupled into two (or more) smaller band complexes. For example, we may write $B=B_1+B_2$, if it decomposes into two complexes $B_1$ and $B_2$. This process does not change the IRRs of the vertices, so we must have $\mathbf c(B)=\mathbf c(B_1)+\mathbf c(B_2)$. Here, $\mathbf c(B_1)$ and $\mathbf c(B_2)$ must be nonzero vectors belonging to $S$. This contradicts our assumption that $\mathbf c(B)\in \text{Hilb}(S)$, since a Hilbert basis vector cannot be generated by other elements of $S$. Therefore, we must have $B$ irreducible.

As for the `only if' part, it is a direct consequence of Proposition 2, which we shall discuss later in Sec.~\ref{VV}.

By this result, the band complexes corresponding to the Hilbert basis vectors can be regarded as elementary building blocks of
a band structure. Such band complexes are referred to as Hilbert band complexes (HBCs). In the following, we illustrate a
procedure to construct HBCs from Hilbert basis vectors of $S$.

We carry out the construction of HBCs in the following two steps. First, pick up a particular Hilbert basis vector $\mathbf v$ on path $U$: $P$-$Q$.
It specifies the multiplicities of $\mathcal{P}_i$ $(i=1,2,\cdots,m)$ and $\mathcal{Q}_j$ $(j=1,2,\cdots,n)$. The sum of the first $m$ entries in $\mathbf v$ gives the number of vertices at point $P$, denoted by $n_P$, namely,
\begin{equation}
  n_P=\sum_{i=1}^m v_i,
\end{equation}
where $v_i$ is the $i$th entry of $\mathbf v$; and the sum of the remaining $n$ entries
in $\mathbf v$ gives the number $n_Q$ of vertices at point $Q$,
\begin{equation}
	\label{eq:198}
  n_Q=\sum_{i=m+1}^{m+n}v_i.
\end{equation}
Separately at $P$ and $Q$, we take all the different energy orderings of the vertices. This gives $n_P! n_Q!/\prod_i v_i!$
configurations.

Second, for each configuration, we connect the vertices by edges to form a HBC.
Each edge corresponds to an IRR $\mathcal{U}_i$ $(i=1,2,\cdots,s)$. Two edges with the same IRR should not cross.
To ensure this condition, we adopt the following approach.
At $P$, we label the vertices by
$P_1$, $P_2$, $\cdots$, $P_{n_P}$, in the order of decreasing energy, similarly for vertices at $Q$.
For each $\mathcal{U}_i$, we identify the number $E_i$ of edges  with IRR $\mathcal{U}_i$ in this HBC from the compatibility relations. Then, we assign the $E_i$ edges among the vertices step by step, according to the following rule: each time we add in an edge connecting a pair of vertices $P_i$ and $Q_j$, we always choose the vertices such that their labels $i$ and $j$ are each the smallest among the available ones. One can easily see that this rule is both sufficient and necessary to ensure that no edges with the same $\mathcal{U}_i$ cross. By going through all $\mathcal{U}_i$ involved in the basis vector, we finish the construction of the HBC.

For example, consider the Hilbert basis vector $\mathbf v^2$ in Eq.~(\ref{15}). Its HBC contains two vertices at $\Gamma$, corresponding to IRRs $\Gamma_1$ and $\Gamma_4$, and two vertices at $X$, both corresponding to the IRR $X_1$. The HBC constructed by our procedure above is plotted in Fig.~\ref{fig:198}(a). Note that for the case with reversed energy ordering of $\Gamma_1$ and $\Gamma_4$, the obtained HBC is just the same as flipping Fig.~\ref{fig:198}(a) upside down. In this work, we will not distinguish such topologically equivalent HBCs.

By using the above approach, we have examined all possible topologically distinct HBCs for paths between high-symmetry points for the 230 space groups. In S4 of Supplemental Material \cite{zhang_supplemental_nodate}, we label the graphs of HBCs by the symbol $H_{m/n}^i$, where $m$ is the number of vertices, $n$ is the number of edges, and $i$ is a third index to distinguish topologically distinct graphs
with the same $m$ and $n$.

We have a few remarks here. First, a Hilbert basis vector may have more than one topologically distinct HBCs, because of different energy ordering of the vertices.
For example, consider the Hilbert basis vector $\textbf{v}^1$ in Eq.~(\ref{15}). It has two distinct HBCs, as shown in Fig.~\ref{fig:198}(b) and \ref{fig:198}(c). The two have different energy ordering of vertices at $\Gamma$ point. By changing the ordering, one can transform one to the other.
This shows that the two HBCs are intrinsically the same thing (corresponding to the same Hilbert basis vector), although they are not equivalent from the graph point of view. 
\begin{figure}[t]
	\includegraphics[width=0.75\linewidth]{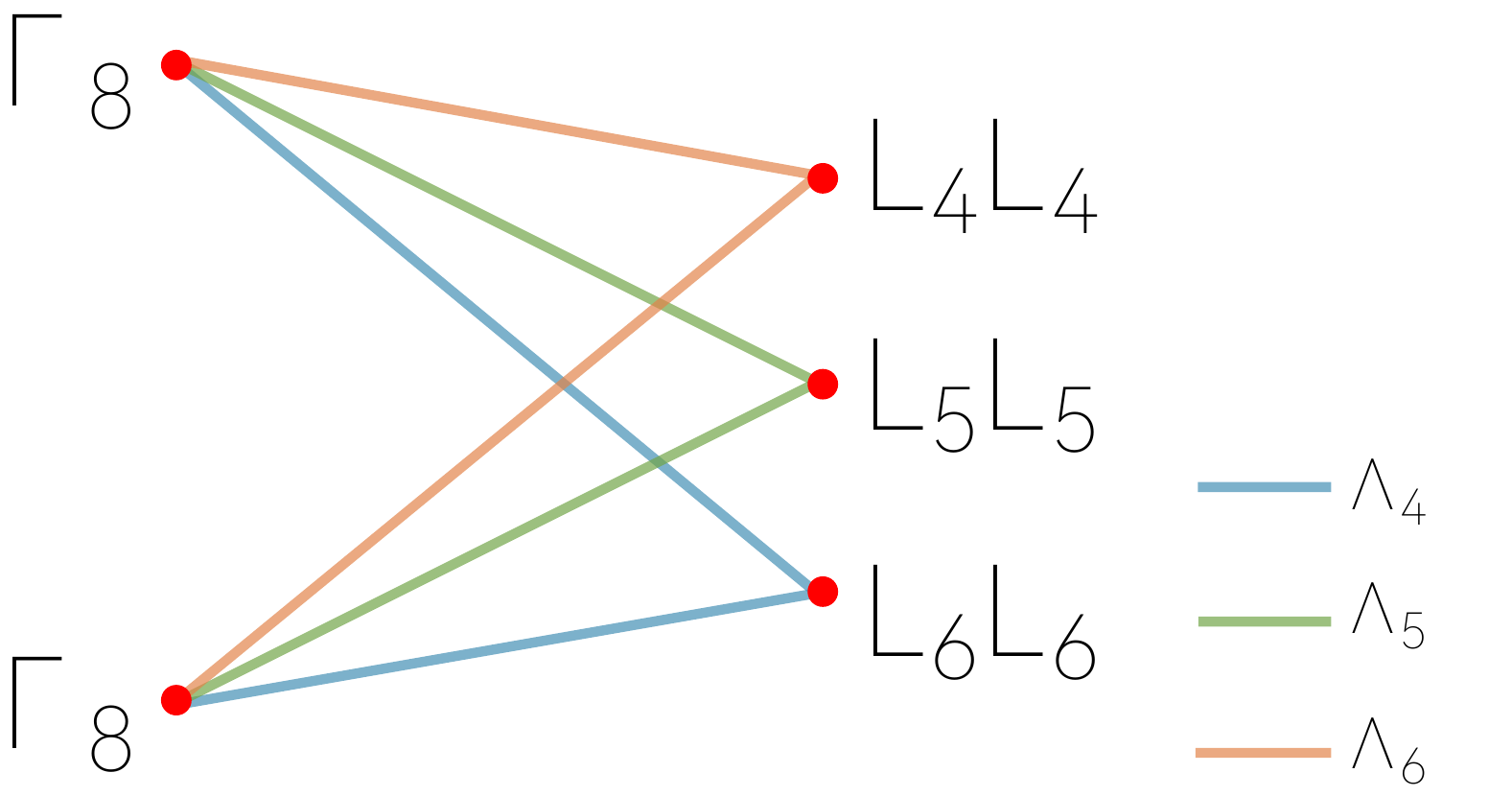}
	\caption{An example Hilbert band complex which is a complete bipartite graph. This complex, labeled as $H_{5/6}^4$, has five vertices and six edges. It can occur on the $\Lambda$ path in space group No.~219.}
	\label{fig:H56}
\end{figure}

Second, viewed as a graph, a band complex can be described by an adjacency matrix \cite{bondy_graph_2008}. It is a square matrix, with entries indicating whether pairs of vertices are adjacent (connected by edges) or not.
Furthermore, such graphs must be bipartite, since an edge can only connect a vertex at $P$ and a vertex at $Q$, it cannot connect two vertices at the same side.
The adjacency matrix for a (undirected) bipartite graph has a symmetric block off-diagonal form. For example, the adjacency matrix for the HBC in Fig.~\ref{fig:198}(a) takes the following form
\begin{equation}\label{18}
  M(H_{4/4}^4)=\left[
                 \begin{array}{ccccc}
                     & \Gamma_1 & \Gamma_4 & X_1 & X_1 \\
                   \Gamma_1 & 0 & 0 & 1 & 0 \\
                   \Gamma_4 & 0 & 0 & 1 & 2 \\
                   X_1 & 1 & 1 & 0 & 0 \\
                   X_1 & 0 & 2 & 0 & 0 \\
                 \end{array}
               \right].
\end{equation}
Here, the number 2 in the entry $(\Gamma_4, X_1)$ corresponds to the two edges connecting the $\Gamma_4$ vertex and the lower $X_1$ vertex in Fig.~\ref{fig:198}(a).

Third, we find that some interesting HBCs may give complete bipartite graphs. A complete bipartite graph is a special bipartite graph where every vertex of the first set (vertices at $P$) is connected to every vertex of the second set (vertices at $Q$). For example, the hourglass band complex in Fig.~\ref{fig:intro}(a) is a complete bipartite graph. We find another  complete bipartite graph, as shown in Fig.~\ref{fig:H56}. This HBC has 5 vertices and 6 edges and is labeled as $H_{5/6}^4$ in our notation. This special complete bipartite HBC may occur in spinless space groups 212, 213, 222, 224, and spinful space group 219.

Fourth, although we may visualize a band complex by a connected graph (which may in turn be described by an adjacency matrix), it must be noted that the graph alone cannot determine whether the band complex is a HBC or not. One must also require the symmetry information, i.e., the IRRs of the vertices or equivalently the $\textbf c$-vector.
For example, consider the $Z$-$A$ path of space group 138. In Ref.~\cite{li_upper_2023}, Li \emph{et al.} found the so-called 8-accordion band complexes on this path, which has the structure shown in Fig.~\ref{fig:acc}(a). 
This complex in Fig.~\ref{fig:acc}(a) is a HBC. Nevertheless, there exists another 8-accordion band complex on this path, as shown in Fig.~\ref{fig:acc}(b), which is not a HBC. This is because by switching its ordering of vertices, one can decompose Fig.~\ref{fig:acc}(b) into two hourglass complexes,
as shown in Fig.~\ref{fig:acc}(c). The difference between Fig.~\ref{fig:acc}(a) and Fig.~\ref{fig:acc}(b) is that they have different $\mathbf c$-vectors. Thus, the symmetry data is needed to fully specify a HBC.

\begin{figure}[t]
	\includegraphics[width=\linewidth]{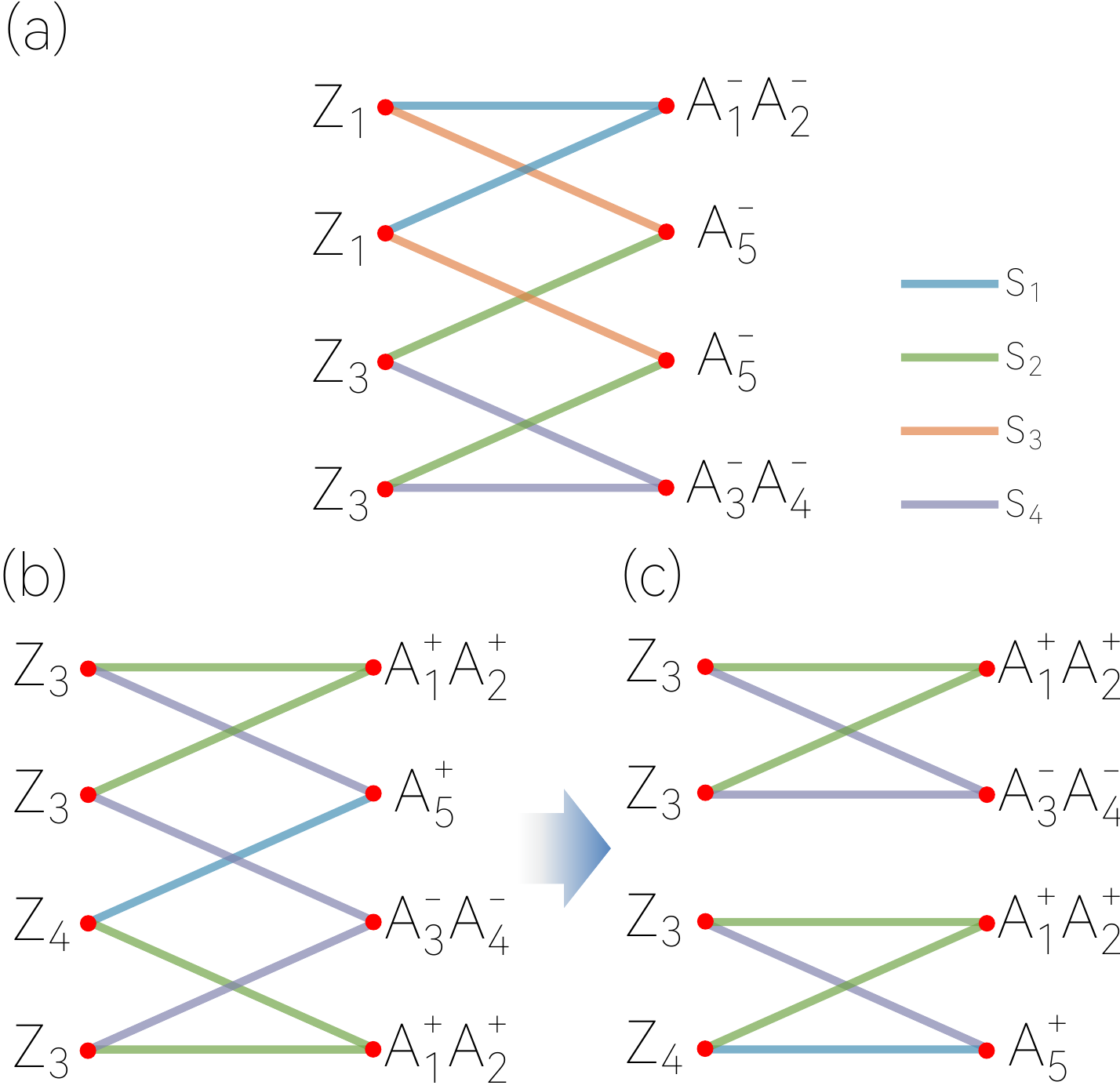}
	\caption{(a) The 8-accordion Hilbert band complex on $Z$-$A$ path of space group No.~138. (b)
Another kind of 8-accordion band complex on the same path. However, this one is not a Hilbert band complex, as it can be decomposed into two hourglass sub-complexes, as shown in (c), by re-arranging the energy ordering of the vertices.
}
	\label{fig:acc}
\end{figure}

Finally, in the above discussion, we focused on a path connecting two high-symmetry points of BZ, but the analysis can be readily extended to multiple paths connecting multiple high-symmetry points. For example, consider the whole BZ of some space group. Suppose there are $p$ high-symmetry points in the BZ. We may analyze all the $p(p-1)/2$ paths between them. The band balance equations (\ref{BB}) can be established for each path, and they can be combined into a large $\mathbf A$ matrix, with the $\mathbf c$ vector
\begin{equation}
  \mathbf c=(c^{\Gamma_1}_1,\cdots,c^{\Gamma_1}_{m_1},c^{\Gamma_2}_{1},\cdots,c^{\Gamma_p}_{m_p})^T,
\end{equation}
where $c^{\Gamma_i}_j$ $(j=1,\cdots,m_i)$ is the number of $j$th IRR at high-symmetry point $\Gamma_i$ that appears in the band structure, and $m_i$ is the number of IRRs at $\Gamma_i$. The length of this vector is $M=\sum_{i=1}^p m_i$.
Then the physical solutions of $\mathbf c$ is the set $S=\text{ker}\mathbf A\cap \mathbb{Z}_{\geq 0}^{M}$, which is again an affine monoid. The Hilbert bases for $S$ can be solved, and the corresponding HBCs can be constructed (which can be done separately for each path), following our procedure introduced above. The HBC here would be a $p$-partite graph, with vertices grouped into $p$ sets according to the $p$ high-symmetry points.

In the following, we shall discuss two applications of HBCs. One is on decomposition of a given band complex. The other is on construction of a large band complex.

\section{decompose a band complex}\label{VV}

As stated in the introduction section, regarding a reducible band complex, a natural question is how to determine its decomposition. This question is answered by the following result.

\textbf{Proposition 2.} A band complex can be decomposed into HBCs, according to how its $\mathbf c$-vector is decomposed by the corresponding Hilbert basis.

Evidently, if a band complex $B$ can be decomposed into a set of HBCs, its $\mathbf c$-vector must be a sum of the
Hilbert basis corresponding to these HBCs. What we need to show is the reverse problem, namely, if we can write
\begin{equation}\label{20}
  \mathbf c(B)=\mathbf v^{i_1}+\mathbf v^{i_2}+\cdots +\mathbf v^{i_q},
\end{equation}
where $i_j$ $(j=1,\cdots,q)$ is an integer between $1$ and {$d=\text{dim Hilb}(S)$}, then $B$ can be decomposed into $q$ HBCs. We may write $B=B_1+B_2+\cdots+B_q$, such that $\mathbf c(B_j)=\mathbf v^{i_j}$. We prove this in the following.

\begin{figure}[t]
	\includegraphics[width=0.8\linewidth]{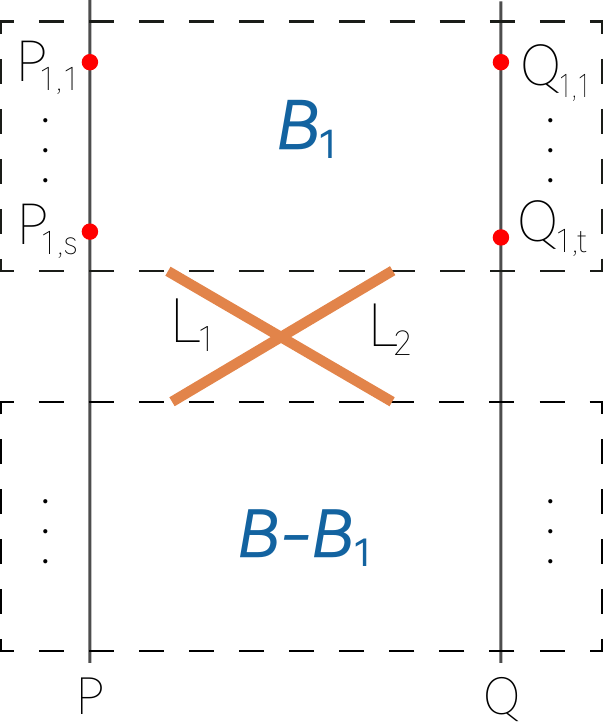}
	\caption{Schematic figure showing the decomposition of a reducible band complex. Please see the description in the main text. }
	\label{fig:proof}
\end{figure}

\emph{Proof.} The above result is trivial if $q=1$, i.e., when $B$ is a HBC. For $q\geq 2$, let's first look at the term
$\mathbf v^{i_1}$ in the decomposition (\ref{20}). This means there is a set of vertices at $P$ and $Q$, labeled as
$\{P_{1,s}\}$ and $\{Q_{1,t}\}$ ($s$ and $t$ are indices labeling the elements of the set), whose IRRs correspond to $\mathbf v^{i_1}$. Note that the choice of the set may not be unique, but it does not affect our argument. Now, to decouple a HBC $B_1$ from $B$, we move the vertices $\{P_{1,s}\}$ and $\{Q_{1,t}\}$ up in energy, so that they are above all remaining vertices in $B$, as illustrated in Fig.~\ref{fig:proof}. We claim that this operation leads to two separate band complexes.
This claim can be proved by contradiction. Suppose that the upper part (corresponding to $\mathbf v^{i_1}$) and the lower part were not disconnected. Then, there must be some vertex, let's say a vertex at $P$, which has an edge $L_1$ connected to the lower part (see Fig.~\ref{fig:proof}). Denote the IRR of this edge by $\mathcal{U}_j$. Now, we note that as a Hilbert basis vector,
$\mathbf v^{i_1}$ satisfies the band balance equations (\ref{BB}). To balance edges with IRR $\mathcal{U}_j$, one can easily see that there must be at least another edge $L_2$, emitted from some vertex at $Q$ and connected to the lower part, as illustrated in Fig.~\ref{fig:proof}. Then, $L_1$ and $L_2$ must cross in between, which is not allowed because they are of the same IRR. This contradiction shows that our operation indeed decouples a complex $B_1$ with $\mathbf c(B_1)=\mathbf v^{i_1}$ from the remaining part of $B$. Repeat this operation for each term in the expansion (\ref{20}), one can decompose a reducible complex into the HBCs, as claimed in the proposition.

As a consequence, if a band complex has its $\mathbf c$-vector not in $\text{Hilb}(S)$, then it must be reducible. This proves the `only if' part of Proposition 1.

We mentioned in Sec.~III that an element of an affine monoid may have more than one way to be generated by Hilbert basis.
By Proposition 2, this indicates that a reducible band complex may have different decompositions into HBCs. We illustrate this by an example. Consider the path $\Lambda$ connecting $\Gamma:(0,0,0)$ and $R:(1/2,1/2,1/2)$ for space group No.~198.
In spinless case, there are three IRRs at $\Gamma$, namely, $\Gamma_1$, $\Gamma_2\Gamma_3$, and $\Gamma_4$, which have been discussed in Sec.~III above Eq.~(\ref{99}). At $R$, there are two 4d IRRs: $R_1R_1$ and $R_2R_3$. On path $\Lambda$, there are three 1d IRRs: $\Lambda_1$, $\Lambda_2$, and $\Lambda_3$ \cite{bradley_mathematical_2009, liu_spacegroupirep_2021, liu_msgcorep_2023}. The $\mathbf A$ matrix in this case is given by
\begin{equation}
  \mathbf A=\left(
              \begin{array}{ccccc}
                1 & 0 & 1 & 0 & -2 \\
                0 & 1 & 1 & -2 & -1 \\
                0 & 1 & 1 & -2 & -1 \\
              \end{array}
            \right).
\end{equation}
The obtained Hilbert basis consists of five elements. They are shown in Table~\ref{tab:1}.

\begin{table}
	\centering
	\caption{List of Hilbert bases on the $\Gamma$-$R$ ($\Lambda$) path of space group No.~198 in the spinless case. The symbols in the top row denote the IRRs at points $\Gamma$ and $R$. }
	\label{tab:1}
	\begin{tabularx}{0.45\textwidth}{@{}Y|YYY|YY@{}}			
	\hline\hline
	&$\Gamma _1$&$\Gamma _2\Gamma _3$&$\Gamma _4$&$R_1R_1$&$R_2R_3$\\
		\hline
	$\mathbf v^{1}$&0&0&4&1&2\\
$\mathbf v^{2}$&0&1&2&1&1\\
$\mathbf v^{3}$&0&2&0&1&0\\
$\mathbf v^{4}$&1&0&1&0&1\\
$\mathbf v^{5}$&2&1&0&0&1\\
		\hline
	\end{tabularx}
	\label{tab:adaptive-table}
\end{table}

\begin{figure}[t]
	\includegraphics[width=0.8\linewidth]{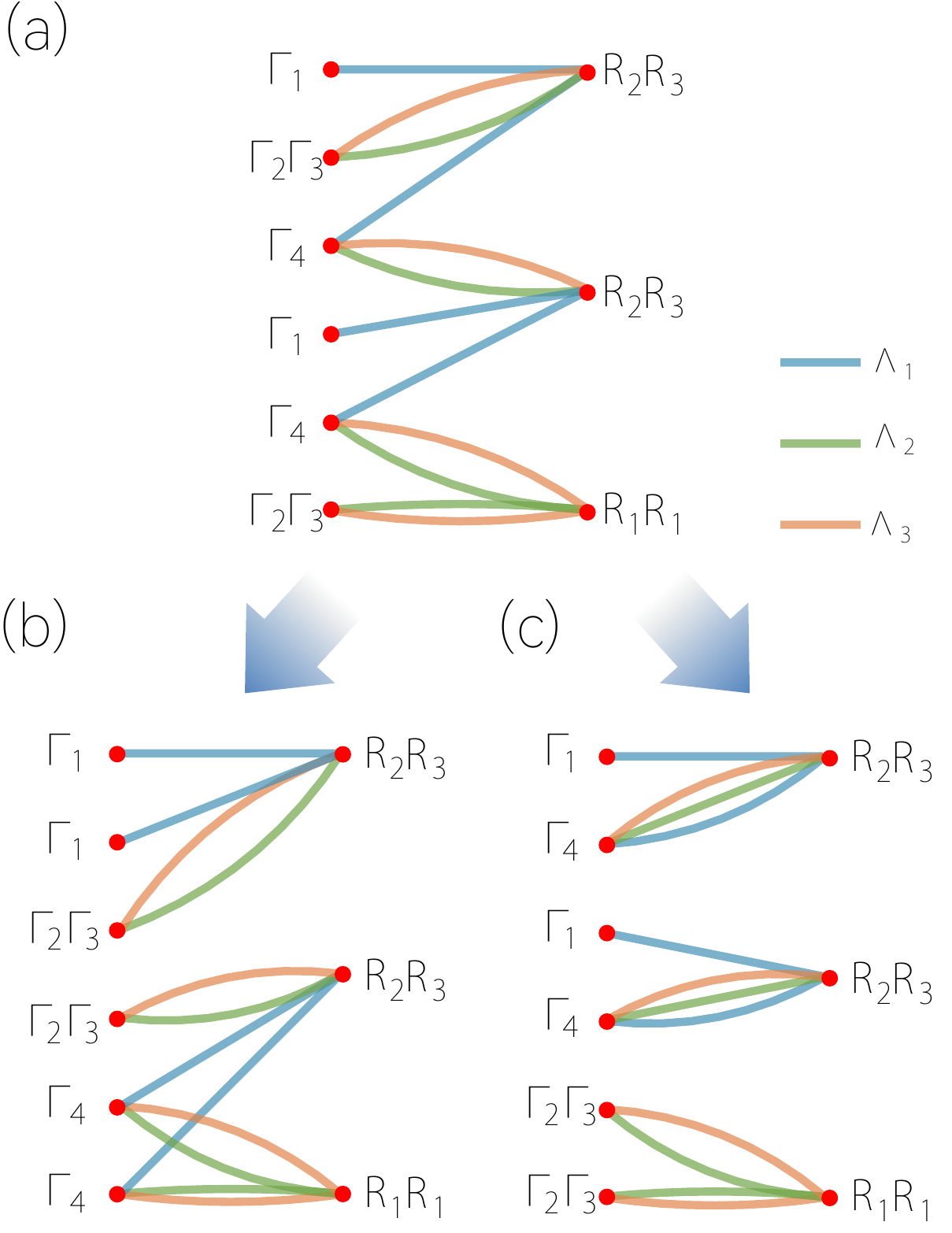}
	\caption{(a) An example band complex corresponding to $\mathbf c(B)$ in Eq.~(\ref{eq:cb}). (b) and (c) show
two different decompositions into Hilbert band complexes, corresponding to Eq.~(\ref{eq:dec}). (b) corresponds to $\mathbf c(B)=\mathbf v^2+\mathbf v^5$, and (c) corresponds to $\mathbf c(B)=\mathbf v^3+\mathbf v^4+\mathbf v^4$.}
	\label{fig:decomposition}
\end{figure}

Let's consider the band complex $B$ in Fig.~\ref{fig:decomposition}(a). Its $\mathbf c$-vector is given by
\begin{equation}
	\label{eq:cb}
  \mathbf c(B)=(2,2,2,1,2)^T.
\end{equation}
From Table~\ref{tab:1}, we note that $\mathbf c(B)$ can have two different decompositions, namely,
\begin{equation}
	\label{eq:dec}
  \mathbf c(B)=\mathbf v^2+\mathbf v^5=\mathbf v^3+\mathbf v^4+\mathbf v^4.
\end{equation}
Thus, the complex $B$ has two different decompositions into HBCs. This is illustrated in Fig.~\ref{fig:decomposition}(b) and \ref{fig:decomposition}(c).

\section{Merge band complexes}

Merging two band complexes to produce a larger complex is the reverse process of the band complex decomposition discussed in Sec.~V. By Proposition 2, one can immediately see that any existing band complex $B$ can be constructed from a set of HBCs, by arranging their vertices according to the prescribed order in $B$.

Moving a step further, we consider the following problem: Given two (or more) HBCs, how to determine whether they can be merged into a larger complex?

Clearly, this merge is not always possible. Consider the simple case where $P$ and $Q$ only have 1d IRRs. Then two HBCs on $P$-$Q$ can never be merged together into a connected piece.

We have not found a simple answer to this problem. Nevertheless, a computational algorithm is straightforward. Given two HBCs, we may form all possible configurations of energy orderings of the vertices, and proceed with the construction process discussed in Sec.~IV for the construction of HBCs. The only difference here is that the constructed ones are no longer guaranteed to be a single complex, i.e., they may be disconnected.  The connectivity of the resulting graph can be determined either by observation or by adopting standard algorithms in graph theory. For instance, one way to check connectivity is to use the
adjacency matrix discussed around Eq.~(\ref{18}). For a $n\times n$ adjacency matrix $M$, one can compute the matrix $W=(1_{n\times n}+M)^n$ \cite{bondy_graph_2008, horvat_igraphm_2023}. A connected graph will have all the entries of this $W$ matrix nonzero.

An interesting result obtained along this line is that we find certain band complex can grow without bound by successively adding a HBC to it. The first example of such unbounded band complex was given in Ref.~\cite{li_upper_2023}, where the basic building blocks are hourglass type HBCs. Below, we present an even simpler example.

\begin{figure}[t]
	\includegraphics[width=\linewidth]{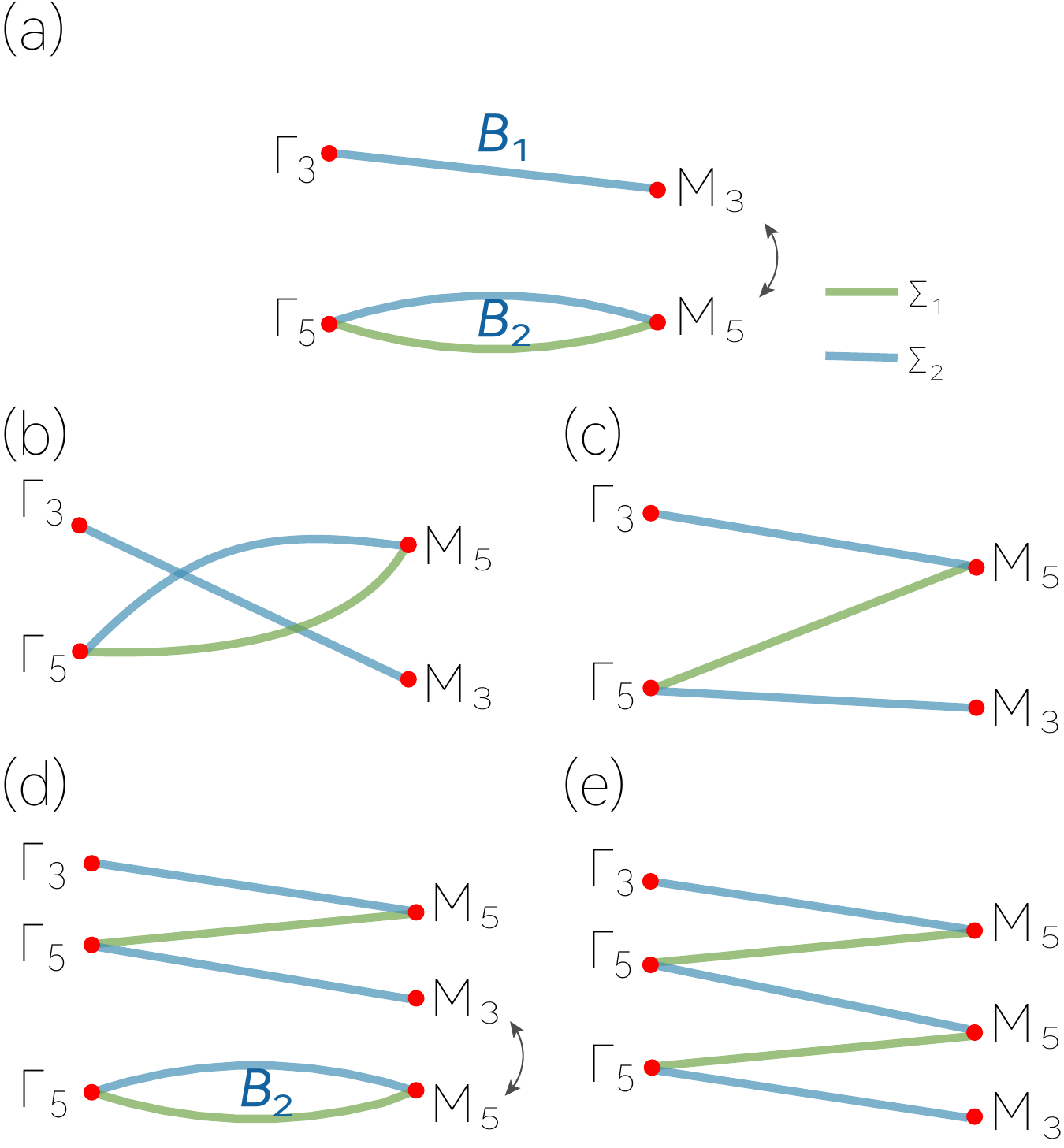}
	\caption{An example showing the construction of an unbounded band complex in space group No.~89. (a) Take two Hilbert band complexes $B_1$ and $B_2$ on the $\Sigma$ path. (b) Merge the two complexes by switching the ordering of vertices $M_3$ and $M_5$. The crossing point marked with the red circle will be gapped and merged complex will become that in (c).
(d-e) We can repeat this process by merging another $B_2$ from below. In this way, the produced zigzag shaped complex can grow without bound. }
	\label{fig:merge}
\end{figure}

Consider the path $\Sigma$ connecting $\Gamma$ and $M$ for space group No.~89. There are 5 IRRs at $\Gamma$ and 5 IRRs at $M$. We will focus on the two HBCs $B_1$ and $B_2$ \cite{bradley_mathematical_2009, liu_spacegroupirep_2021, liu_msgcorep_2023}, as illustrated in Fig.~\ref{fig:merge}(a). The IRRs of the vertices and the edges are indicated in the figure.
In Fig.~\ref{fig:merge}(a), HBC $B_2$ is arranged below $B_1$. We can merge them by switching the energy ordering of the two vertices $M_3$ and $M_5$ (see Fig.~\ref{fig:merge}(b)). This leads to the zigzag shaped band complex in Fig.~\ref{fig:merge}(c). Interestingly, the lower part of this resulting complex is the same as that of $B_1$, so one can continue to merge a $B_2$ to it (see Fig.~\ref{fig:merge}(d) and \ref{fig:merge}(e)). This process can keep going on indefinitely, resulting in a band complex with $2n+1$ edges, with $n\in\mathbb{Z}_{\geq 0}$ arbitrarily large.

\section{discussion and conclusion}

We have shown that HBCs are a useful tool in analyzing band complexes. Using the method developed here, we have
obtained all non-isomorphic HBCs (from the graph perspective) on a path between any two high-symmetry points in the 230 gray space groups. We find there are 133 different kinds of graphs, in which 119 (36) can occur in spinless (spinful) systems.
The largest HBC contains 13 vertices and 16 edges, which occurs in spinless (single-valued) space groups 220 and 230.
As for the spinful case, the largest HBC contains 12 vertices and 12 edges, which appears in space groups 169, 170, 178 and 179. More details can be found in S5-S7 of Supplemental Material \cite{zhang_supplemental_nodate}.

As discussed at the end of Sec.~IV, it is straightforward to consider the global Hilbert bases defined for the whole BZ and
the associated global HBCs which are multi-partite graphs. The decomposition of a global band complex and the merge of global HBCs follow the same way as discussed in Sec.~V and Sec.~VI. In addition to these applications, the global Hilbert bases, which we denote as GHB, are also useful in diagnosing topological insulator states.
For example, a class of quantum band insulator (QBI) states \cite{po_symmetry-based_2017} correspond to the following set difference
\begin{equation}
  \{\text{QBI}\}=\text{Aff}\langle \text{GHB}\rangle_{\mathbb{Z}_{\geq 0}}-\text{Aff}\langle \text{EBR}\rangle_{\mathbb{Z}_{\geq 0}},
\end{equation}
where $\text{Aff}\langle \text{GHB}\rangle_{\mathbb{Z}_{\geq 0}}$ is the affine monoid generated by the GHBs
(with nonnegative coefficients), and EBR is the so-called elementary band representations, which are the minimal sets of IRRs
that can be realized by atomic models \cite{zak_band_1982, michel_connectivity_1999, michel_elementary_2000}. In addition, a class of fragile topological insulators (FTIs) \cite{po_fragile_2018} can also be characterized with the help of GHBs:
\begin{equation}
  \{\text{FTI}\}=\text{Span}\langle \text{EBR}\rangle_{\mathbb{Z}}\cap \text{Aff}\langle \text{GHB}\rangle_{\mathbb{Z}_{\geq 0}}-\text{Aff}\langle \text{EBR}\rangle_{\mathbb{Z}_{\geq 0}},
\end{equation}
where $\text{Span}\langle \text{EBR}\rangle_{\mathbb{Z}}$ is the linear space spanned by EBR (with coefficients in $\mathbb{Z}$). In a recent work, Song \emph{et al.} \cite{song_fragile_2020} have utilized GHBs to develop a systematic approach for diagnosing fragile topological insulators.

In conclusion, we have developed the concept of HBCs and the relevant method for studying band connectivity. These new tools help to solve fundamental questions regarding reducibility, decomposition, and merge of band complexes. We show that the reducibility of a band complex is completely characterized by its $\textbf{c}$-vector, i.e., whether the vector is a Hilbert basis vector or not. The decomposition of a reducible band complex is also completely determined by how its  $\textbf{c}$-vector is decomposed by Hilbert basis vector, although the decomposition may not be unique.  
We develop algorithms to construct HBCs for each Hilbert basis vector, analyze HBCs from graph perspective, and build larger band complexes by merging HBCs. We find certain unbounded band complexes, which can grow indefinitely by successfully merging a HBC to it. These findings will deepen our fundamental understanding of band theory and offer us new methods for band structure analysis.

\begin{acknowledgements}
  The authors thank D. L. Deng for valuable discussions. We acknowledge support from the NSF of China (Grants Nos.  12234003, 12321004, 12347214, 12004028), the Fundamental Research Funds for the Central Universities (No. ZY2418), and UM start-up grant (SRG2023-00057-IAPME). 
\end{acknowledgements}

\bibliography{Hilbert.bib}
\bibliographystyle{apsrev4-2}

\end{document}